\documentclass[11pt,preprint]{aastex}
\usepackage{epsfig,color}
\usepackage{mathrsfs}
\usepackage{ulem}
\usepackage{amsmath}
\usepackage{empheq}
\usepackage{bm}
\usepackage{lscape}

\def\OXYZ{O\!-\!\!XY\!Z}

\def\bhm{M_{\bullet}}

\def\RR{\vec{R}}

\def\nn{\vec{n}}

\def\Rin{R_{\rm in}}
\def\Rmid{R_{\rm mid}}
\def\Rout{R_{\rm out}}

\def\Rdi{R_{1, {\rm d}_i}}
\def\Rini{R_{1, {\rm in}_i}}
\def\Routi{R_{1, {\rm out}_i}}
\def\Gw{\Gamma_{\!\omega}}

\def\etal{{et al.}}

\def\ergs{{\rm erg~s^{-1}}}

\def\kms{\rm km\,s^{-1}}

\def\sunm{M_{\odot}}

\def\oiii{[O~{\sc iii}]}

\begin{document}

\title{Kinematic signatures of reverberation mapping of 
close binaries of supermassive black holes in active galactic nuclei}

\author{
Jian-Min Wang\altaffilmark{1,2,3},
Yu-Yang Songsheng\altaffilmark{1,2},
Yan-Rong Li\altaffilmark{1} and
Zhe Yu\altaffilmark{1,2}\\ ~\\
{accepted by {\it The Astrophysical Journal}}
}

\altaffiltext{1}
{Key Laboratory for Particle Astrophysics, Institute of High Energy Physics, CAS, 19B Yuquan Road,
Beijing 100049, China, wangjm@ihep.ac.cn}

\altaffiltext{2}
{School of Astronomy and Space Science, University of Chinese Academy of Sciences, 19A Yuquan Road, 
Beijing 100049, China}

\altaffiltext{3}
{National Astronomical Observatories of China, CAS, 20A Datun Road, Beijing 100020, China}

\begin{abstract}
Close binaries of supermassive black holes (CB-SMBHs) with separations of $\lesssim 0.1$pc as the final stage of galaxy mergers are sources of low frequency gravitational waves (GW), however, they are still elusive observationally because they are not spatially resolved. Fortunately, reverberation as echoes of broad emission lines to ionizing continuum conveys invaluable information of the dynamics of broad-line regions (BLRs) governed by supermassive black holes in the central regions of active galactic nuclei (AGNs). In this paper, we demonstrate how to composite the hybrid 2-dimensional transfer functions of binary BLRs around the CB-SMBHs in AGNs, providing an opportunity of identifying them from reverberation mapping (RM) data. It is found that there are variation-coupling effects in the transfer functions, arising from the coupling of CB-SMBH light curves in the Fourier space. We provide semi-analytical formulations of the transfer functions for kinematic maps of the gas. For cases with the simplest variation-coupling effects, we make calculations for several BLR models and reveal significant distinctions from those of single active black holes. In principle, the difference is caused by the orbital motion of the CB-SMBH systems. In order to search for CB-SMBHs in time-domain space, selection of target candidates should focus on local AGNs with H$\beta$ double-peaked profiles and weaker near-infrared emission. High-fidelity RM-campaigns of monitoring the targets in future will provide opportunities to reveal these kinematic signatures of the CB-SMBHs and hence for measurements of their orbital parameters.
\end{abstract}


\section{Introduction}
As a milestone in natural science, LIGO's exciting discoveries of 
$\sim 10^2\,$Hz GW from stellar-mass black hole binaries (Abbott et al. 2016; 2017a,b)
greatly advanced the understanding of general relativity. It is eagerly desirable  for physicists 
and astronomers to detect low-frequency gravitational waves (GWs) of supermassive black hole (SMBH) 
binaries (e.g., Sesana 2013; Shannon et al. 2015;
Mingarelli et al. 2017; Sesana et al. 2017) for many years. SMBH binaries with separations of $\sim$ kpc 
have been found and are quite common (Komossa et al. 2003; Bianchi et al. 2008; Wang et al. 2009;
Comerford et al. 2009, 2013, 
2015; Green et al. 2010; Koss et al. 2011; Liu et al. 2011, 2017; Fu et al. 2015) as a natural consequence 
of galaxy mergers (Begelman et al. 1980; Haelnelt \& Kauffmann 2002; Volonteri et al. 2003; Merritt 
\& Milosavljevi\'c 2005; Colpi \& Dotti 2011; Rasskazov \& Merrittt 2017), but close binaries of SMBHs 
(CB-SMBHs), which are bounded by the gravity of the double SMBHs, are still unknown so far because of 
lack of robust criteria to observationally identify these spatially unresolved sources. 

Theoretical features of CB-SMBHs have been predicted for observations, however, observational 
data can be alternatively
explained by complicated BLR models around a single black hole. Thus, they are still elusive though
there is growing evidence for appearance of CB-SMBHs. As one of the several signatures of CB-SMBHs, 
radial velocity curves of
double-peaked H$\beta$ emission line (Popovi\'c et al. 2000; Tsalmantza et al. 2011; Popovi\'c 2012; 
Shen \& Loeb 2010), but the asymmetry is a better tracer than the double-peakedness in
reverberations (Shen \& Loeb 2010). Candidates with this feature are 
NGC 4151 (Bon et al. 2012) and NGC 5548 (Li et al. 2016), which show opposite motions of
red and blue peaks. A large sample of AGNs with double-peaked H$\beta$ profiles built up by
Eracleous et al. (2012) have been monitored for systematic shifts of the two peaks for a couple of
years (Runnoe et al. 2017). Periodical variations 
could be regarded as a signature of binary black holes, such as 
OJ 287 (Valtonen et al. 2008), PG 1302-102 (Graham et al. 2015), Akn 120 (Li et al. 2017) and
others (Liu et al. 2015; Zheng et al. 2016; Dorn-Wallenstein et al. 2017). 
It usually takes at least three times of the periods to determine periodicity of varying AGNs,
such as,  more than 10 years to justify their periodicity if the orbital periods are 3 years 
or so (Li \& Wang 2018). X-shaped radio jet has been suggested as a signature of binary black holes 
(Merritt \& Ekers 2002; Cheung 2007), but only one could be plausible (Kharb et al. 2017).
Deficit of UV emissions of spectral energy distributions is formed by the interaction between the
secondary and the circumbinary disks (Hayasaki et al. 2008; Schnittman 2011, 2013; Sesana et
al. 2012; Roedig et al. 2014). However, there are alternative explanations as to these phenomena, 
such as, processing jet as alternative explanations of AGN periodical variabilities  
(since most of them
are radio-loud AGN), dust extinctions for UV deficit in Mrk 231 (Yan et al. 2016; but see Leighly 
et al. 2016), elliptical disks (Eracleous et al. 1995), hot spots (e.g., Jovanovi\'c et al. 2010) 
or spiral-arm (e.g., Storchi-Bergmann et al. 2017) for double-peaked profiles of broad H$\beta$ line. 
Three candidates with  the double-peaked profiles, 
Arp 102B, 3C 390.3 and 3C 332 have been excluded by long-term monitoring campaigns (Eracleous et al. 
1997). No CB-SMBHs have been unambiguously identified so far.

On the other hand, background of low-frequency GWs as an assembly generated by CB-SMBH
populations in the Universe have been explored for many years, but it has been not detected 
successfully, yet (e.g., Shannon et al. 2015). Recently, it has been realised that 
CB-SMBH identifications are very important for GW detection of individual binaries (Rosado et 
al. 2015; Rosado et al. 2016; Wang \& Mohanty 2017). In such a case of an individual CB-SMBH, 
it will be necessary for optical monitoring campaigns to measure its orbital parameters, such 
as its SMBH mass, mass ratio and separations, and to perform measurements of nano-Hertz GWs. It 
is almost for sure that this can be only done by RM-campaigns. It is an urgent task 
for astronomers to efficiently identify CB-SMBHs in campaigns much shorter than the
CB-SMBH orbital periods (usually at a level of 10 years) and measure their orbital parameters.

Reverberation mapping (RM) is a powerful tool to probe the kinematics of the central regions of 
AGNs (Blandford \& Mckee 1982; Peterson 1993). In particular, the velocity fields of the ionized 
gas in the broad-line regions (BLRs) can be re-constructed by the maximum entropy method from RM 
data (e.g., Horne 1994) and hence the central potentials. In such a system containing 
a CB-SMBH, orbital motion is the key features of the BLR kinematics, which is 
expected to reveal CB-SMBHs in campaigns much shorter than
their orbital periods. 

\section{Composite responses from binary BLRs}
RM technique has been discussed only in the frame of single black hole so far, by 
theory (or simulations), or for explanations of RM data. Pioneering work on reverberation 
of a single BLR of a single black hole was promoted by Blandford \& McKee (1982; hereafter BM82).
As an important description of the BLR, transfer functions (TFs) delivering its kinematics and 
structure information can be derived directly from observational data for comparisons with 
theoretical models. We extend the pioneering work to the case of binary BLRs. 

Generally, given the light curves of an emission line and continuum, we have 
\begin{equation}
L_{\ell}(v,t)=\int_{-\infty}^{\infty}dt^{\prime}L_{\rm c}(t^{\prime})\Psi(v,t-t^{\prime}),
\end{equation}
where $\Psi(v,t)$ is the 2-dimension TF, $\Psi(v,t)=0$ for $t<0$ and $\Psi(v,t)\ge 0$ for 
$t\ge0$, and the subscripts $\ell$ and c
represent line and continuum, respectively. It has been demonstrated by BM82 that
\begin{equation}
\Psi(v,t)=\frac{1}{2\pi}\mathscr{F}^{-1}\left[
          \frac{\tilde{L}_{\ell}(v,\omega)}{\tilde{L}_{\rm c}(v,\omega)}\right],
\end{equation}
where 
\begin{equation}
\tilde{L}_{\ell,\rm c}=\mathscr{F}\left[L_{\ell,\rm c}(v,t)\right],
\end{equation}
$\omega$ is frequency, ($\mathscr{F},\mathscr{F}^{-1}$) are the Fourier and the inverse 
Fourier transform of the light curves, respectively.

For a CB-SMBH system, we suppose that there are binary BLRs, but they are independently 
photoionized only by accretion disks of their own black holes (their circumbinary disk 
is neglected mostly emitting optical photons). We denote this  
detached CB-SMBHs. Two ionizing sources are independently varying and can be described by 
$L_{1,2}^{\rm c}(t)$ for the continuum, and lead to $L_{1,2}^{\ell}(v,t)$  of the broad
emission lines. Since CB-SMBHs are usually spatially unresolved, we only receive 
the total fluxes of emission lines and continuum.
According to Equation (2), the total TF of binary BLRs can be expressed by 
\begin{equation}
\Psi_{\rm tot}(v,t)=\frac{1}{2\pi}{\mathscr{F}}^{-1}
          \left[\frac{\tilde{L}_1^{\ell}(\omega)+\tilde{L}_2^{\ell}(\omega)}
          {\tilde{L}_1^{\rm c}(\omega)+\tilde{L}_2^{\rm c}(\omega)}\right]
          =\frac{1}{2\pi}{\mathscr{F}}^{-1}
                    \left[\frac{{\cal L}_1(\omega)}{1+\Gw}
                    +\frac{{\cal L}_2(\omega)}{1+\Gw^{-1}}\right],
\end{equation}
where ${\cal L}_{1,2}(\omega)=\tilde{L}_{1,2}^{\ell}(\omega)/\tilde{L}_{1,2}^{\rm c}(\omega)$,
and $\Gw=\tilde{L}_2^{\rm c}(\omega)/\tilde{L}_1^{\rm c}(\omega)$ indicates a couple of 
continuum variations in the frequency space arisen from the Fourier transformations. 
We call this as variation-coupling effect and $\Gw$ is
a key parameter determined by the properties of the continua associated with each of the binary black holes.
Given the emissivity law [$\epsilon(\RR)$] based on the geometries of the BLR and the projected velocity
distribution [$g(\RR,v)$], we have
\begin{equation}\label{psitot}
\Psi_{\rm tot}(v,t)=\sum_{k=1}^2\int d\vec{R}_k\,{\cal H}_k(v,\RR_k)
                    {\cal{Q}}_k(v,t,R_k),
\end{equation}
where ${\cal H}_k(v,\RR_k)=\epsilon(\RR_k)g(\RR_k,v)/4\pi R_k^2$, and 
\begin{equation}
{{\cal{Q}}_{1,2}(v,t,R_{1,2})}=\frac{1}{2\pi}\int_{-\infty}^{+\infty}\!\!d\omega
      \left[\frac{e^{-i\omega t_1}}{1+\Gw},\frac{e^{-i\omega t_2}}{1+\Gw^{-1}}\right],
\end{equation}
where the subscripts correspond to the first and the second terms in the mid-bracket, 
respectively, and  $t_{1,2}=t-(R_{1,2}+\vec{R}_{1,2}\cdot\nn_{\rm obs})/c$, $\nn_{\rm obs}$ 
is the observer line of sight and $c$ is the light speed. 

In a single BLR, ${\cal{Q}}_{1,2}\equiv\delta(t^{\prime})/2$ with
$t^{\prime}=t-(R+\vec{R}\cdot\nn_{\rm obs})/c$ leads to a simple 
expression (Blandford \& Mckee 1982), where $\delta(t^{\prime})$ is the $\delta$-function. 
We  stress that the total TF is not a simple summation of two 
individual functions as shown by Eq. (\ref{psitot},\,\ref{psitot0}) due to ${\cal Q}_{1,2}$.
Actually, ${\cal Q}_{1,2}$ indicates a couple of the AGN continuum variability patterns 
and the BLR geometries, arising from the spatially unresolved effect of the binary black 
holes.

In principle, $\Gw$ is not well known from accretion-disk theories. Fortunately, the observational
properties of long-term variations of the optical continuum provide clues to $\Gw$ from several 
large campaigns of monitoring AGNs. Early studies of AGN long-term variations show
that the power-density spectra can be characterized by a 
power-law as ${\rm PSD}(\omega)\propto \omega^{-\gamma_{\omega}}$ with a very low-frequency break, 
$\gamma_{\omega}\sim 2$ in
PG quasars (Giveon et al. 1999) which were targets of reverberation mapping campaign (Kaspi et 
al. 2000); $\gamma_{\omega}=2.13^{+0.22}_{-0.06}$ in 13 AGNs (Collier 
\& Peterson 2001); $\gamma_{\omega}=1.77$ in MACHOS quasars (Hawkins 2007). It
should be noted that $\gamma_{\omega}=2$ corresponds to
a continuous time first-order autoregressive process (Kelly et al. 2009).
Recent data of {\it Kepler} observations of 21 AGNs show $\gamma_{\omega}=1.75-3.2$ deviating from
$\gamma_{\omega}=2$, but the $\gamma_{\omega}-$distribution peaks around $\gamma_{\omega}\approx 2.5$ 
(Mushotzky et al. 2011; Smith et al. 2018). This indicates that AGN variations deviate from the 
damped random walk models, but also implies that $\Gw$ as a ratio of two PSDs could be $\omega$-free
beyond break frequencies of the PSDs in light of the $\gamma_{\omega}-$distribution. 
In principle, Equation (6) can be expressed semi-analytically, however, its expression is
formidable due to $\Gw$, and  $\Gw-$effects are briefly discussed in Appendix A.

The simplest case is that the double black holes have the same properties of continuum variations,
i.e. they have approximately same PSD. In such a case, we have $\Gw\approx \Gamma_0={\rm constant}$, 
yielding the following formulations from Eq. (5)
\begin{equation}\label{psitot0}
\Psi_{\rm tot}(v,t)=\frac{\Psi_1(v,t)}{1+\Gamma_0}+
                    \frac{\Psi_2(v,t)}{1+\Gamma_0^{-1}},
\end{equation}
where  
\begin{equation}
\Psi_{1,2}(v,t)=\int d\vec{R}_{1,2}\,{\cal H}_{1,2}(v,\RR_{1,2})\delta(t_{1,2}),
\end{equation}
are the 2D-TFs of each AGN analytically expressed later. The parameter $\Gamma_0$ cannot be
$\Gamma_0\gg1$ or $\Gamma_0\ll1$, avoiding one of the binary BLRs dominates over another. The 
present scheme for CB-SMBHs is only valid for $\Gamma_0\sim 1$, likely for high mass ratio systems.

Here two distinguished features of $\Psi_{\rm tot}$ are stressed: 1) coupling effects ($\Gw$ 
or $\Gamma_0$) in time domain; 
2) the orbital motions included in the total function. Orbital motions are the major driver
of kinematics distinguishing from the case of a single black hole. The
coupling coefficient $\Gw$ results in a more complicated composition of 2D-TFs, but it could be
a free parameter in light of $\gamma_{\omega}$-peaked distribution. We may take $\Gw=\Gamma_0$ 
as a free constant when comparing the theoretical models with observational data through Markov 
Chain Monte-Carlo (MCMC) simulations for orbital parameters.
The subsequent sections are devoted to calculate 2D-TF for given geometries and dynamics.

\section{Two-dimensional transfer functions}
\subsection{Geometries}
All discussions in this paper are based on the broad H$\beta$ line, which is the most popularly 
monitored in RM campaigns. In light of the high-quality velocity-resolved RM, a single BLR is 
commonly composed of either a geometrically thin/flattened disk, or inflows, or very few  
outflows (e.g., Grier et al. 2009, 2012, 2013; 
Denney et al. 2009, 2011; Bentz et al. 2010, 2012; Du et al. 2016) for about 15 AGNs or so. 
These geometries are employed in modelling 
the BLR for accurate measurements of black hole masses (Pancoast et al. 2011, 2014; Li et al. 2013,
Grier et al. 2017). A brief review on BLR models can be found in Goad \& Korista (2012, 
also for an extensive list of references, or in Wang et al. 2017) as well as for a brief
comparison with observations. Some suggestions of BLR originations linked with torus have been made
by Goad \& Korista (2012). In particular, these ingredients of the BLRs may originate 
from the tidally disrupted clumps in duty torus (Wang et al. 2017). In this paper, we assume 
that CB-SMBH BLRs are composed of two independent BLRs, and each BLR can be described by either 
disk-like with Keplerian rotation (i.e., virialized part of the BLR), or inflows or outflows. 

\begin{figure*}[t!]
\begin{center}
\includegraphics[angle=0,origin=c,trim=150pt 10pt 90pt 80pt,width=0.38\textwidth]{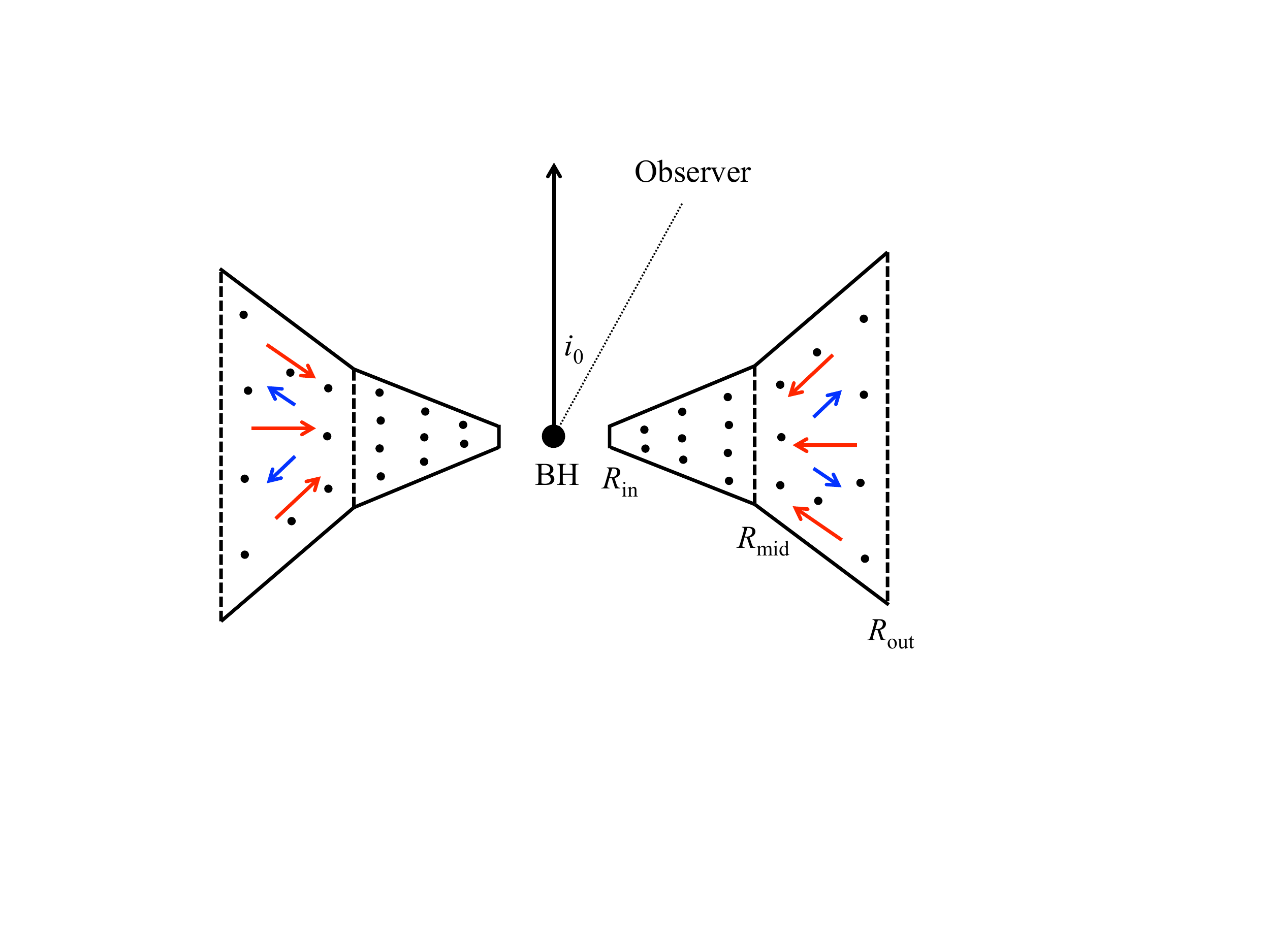}
\includegraphics[angle=0,origin=c,trim=90pt -90pt 90pt 80pt,width=0.36\textwidth]{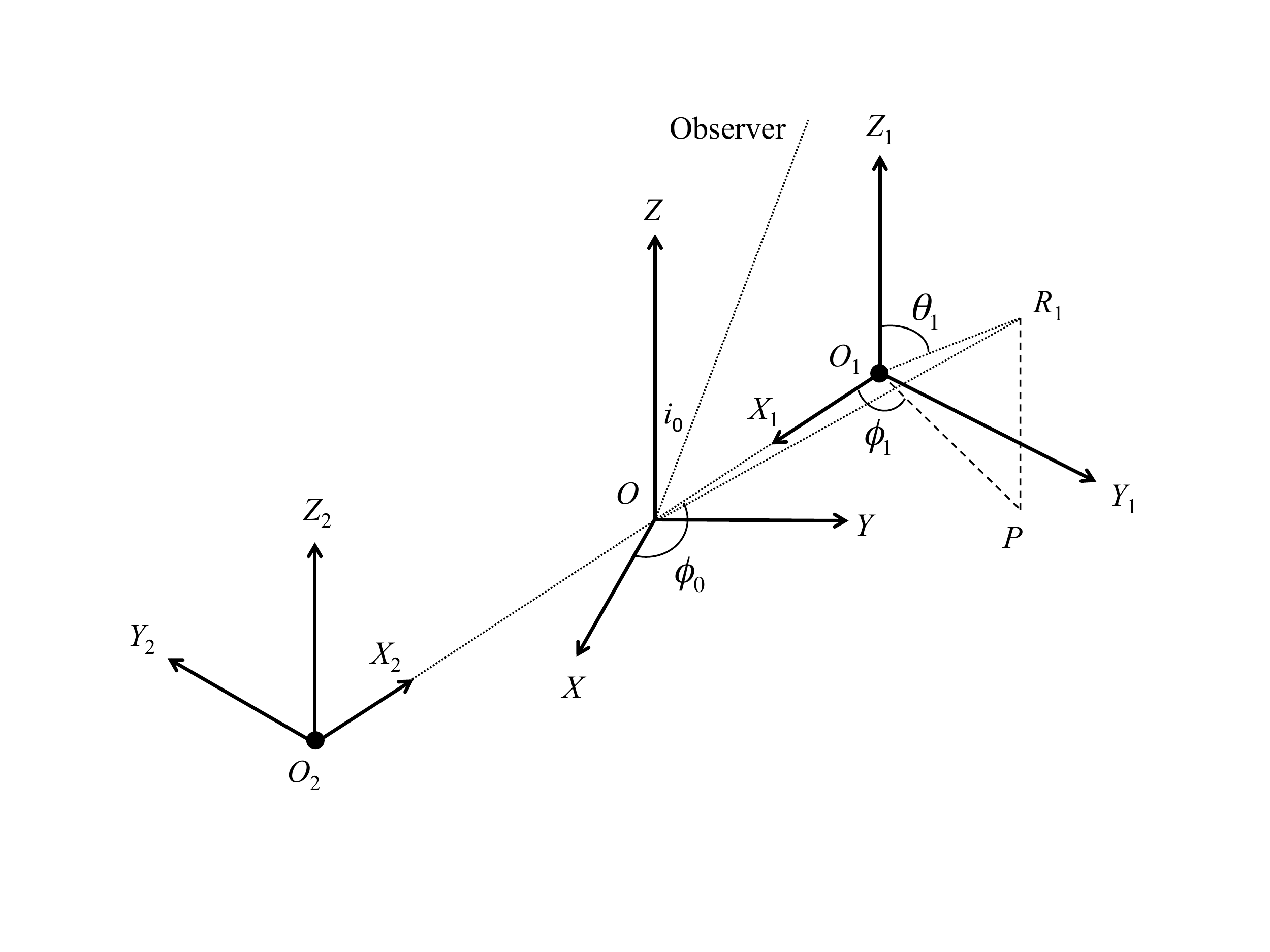}
\end{center}
\vglue -2.2cm
\caption{\footnotesize {\it Left}: Cartoon model of the BLR of one individual 
black hole. The flattened disk part between $\Rin$ and $\Rmid$ is Kepler-rotation dominated, and the 
inflows and outflows between $\Rmid$ and $\Rout$
are non-virialized (but they are assumed to be co-spatial for simplicity). The
relative strengths of the three components determine the profiles of broad emission lines.  
{\it Right}: Geometric relations of CB-SMBHs with two BLRs. For simplicity, we assume
that the two BLRs and their orbital motions are co-planar and they have parallel rotation axes with the
orbit of the binary system. $O$ is the mass center of the binary black holes,
and the observer is in the $OYZ-$plane. $O_1$ and $O_2$ are the primary and secondary black holes with
a distance $|O_1O_2|=A_0$,
and $O_1X_1$ and $O_2X_2$ are toward to the mass center, and $O_1Y_1$ and $O_2Y_2$ are anti-parallel. 
Anticlockwise is positive. $i_0$ is the inclination angle with respect to the orbital plane.
$\phi_0$ is the phase angle of the binary orbit at the observational epoch. Such a geometric relation 
can be obtained by parallel shifting $\OXYZ$ by a distance $A_1$ from $O$ to $O_1$ and rotating $O_1X_1Y_1$
around $O_1Z_1$ by $\pi+\phi_0$. $\vec{R}_1$ is any point of the BLR-1 and for $(\OXYZ)_2$ accordingly.
}
\label{fig:1}
\vglue -0.4cm
\end{figure*}

Figure 1 panel (left) shows the geometry of a single BLR with innermost and outer radii ($\Rin$ 
and $\Rout$, respectively). The flattened disk part is the innermost part of the BLR, which is 
evidenced by velocity-resolved delay maps. Their red and blue wings indicate disk-like 
geometry, for example NGC 5548 (Denney et al. 2009, 2010; Grier et al. 2013;
Bentz et al. 2010; Lu et al. 2016). We use the radius 
$\Rmid$ to describe a transition from disks to inflows or outflows in a single BLR. 
If $\Rin=\Rmid$, the BLR is purely composed of inflows. When $R_{\rm mid}=R_{\rm out}$, the BLR is
a pure disk. Though outflows are not so common in H$\beta$ reverberation, we still cover it by
including outflows co-spaced with the inflows for simplicity. In principle, outflows could be a 
totally free component, but its effects can be found from the same procedures. 

For each BLRs, we assume that they follow the canonic $R-L$ relation (Kaspi et al. 2000; 
Bentz et al. 2013, but see Du et al. 2015, 2018), and the binary BLRs are separated. This provides 
spatial limits on the BLRs of the binary system. We also assume
that each BLR is only photoionized by its own accretion disk of black holes. 
In principle, the present scheme is easily extended to the merged 
BLRs commonly shared by the binary accreting black holes (Songsheng \& Wang 2018, in preparation). 
The inner radius of the circumbinary disk is about the Hill radius 
$R_{\rm H}\approx 0.7\,q^{1/3}A_0$, where $A_0$ is the separation between the two black holes and
$q$ is their mass ratio (Artymowicz \& Lubow 1996; Hayasaki et al. 2008; Cuadra et al. 2009;
D'Orazio et al. 2013; Farris et al. 2014), but the black holes are accreting from the circumbinary
disk. In such a scenario, the BLRs are smaller than the separation. 
The circumbinary disk is only emitting optical photons which
are not ionizing the BLRs.

We would remind readers of that BLR responses depend on spatial emissivity 
which is determined by spatial distribution of clouds in BLRs and spectral energy distribution 
of ionizing sources (see detailed calculations made by Goad et al. 2012; Goad \& Korista 2014, 2015). 
Therefore the observed echoes of the broad H$\beta$ line are emissivity weighted responses, and the 
transfer functions are 
emissivity weighted then as well as for the cases of binary black holes. In this paper, we simply assume
that the responsivity is linear proportional the distributions of number density of cloud of the BLRs.

\subsection{Kinematics of the ionized gas}
In the local frames, the disk-component of the BLRs is assumed to have the Keplerian rotation velocity 
$|\vec{V}_{\rm d}|=V_{\rm K}$. This simplified assumption is supported by the correlation of
${\rm FWHM}\propto \tau^{-1/2}$ between emission line lags ($\tau$) and the full-width-half-maximum 
(FWHM) in several well-mapped AGNs, such as, NGC 5548
showing virialized stratified structures from high-ionisation lines to low-ionisation lines (Peterson \& 
Wandel 1999), 3C 390.3 and NGC 7469 (Peterson \& Wandel 2000, and see more data in recent references). 
For the inflows/outflows, we assume that
\begin{equation}
V_{\rm in,out}=\left\{\begin{array}{l}
\displaystyle{\alpha_0\left(\Rout/R\right)^{\alpha}V_{\rm K}},\quad ({\rm inflows}),\\
   \\
\displaystyle{-\beta_0\left(R/\Rout\right)^{\beta}V_{\rm K}},\quad ({\rm outflows}).\\
\end{array}\right.
\end{equation}
The inflow velocity is generally slower than the free-fall whereas the outflow has velocities faster
than the escaping velocity. The parameters $(\alpha,\alpha_0,\beta,\beta_0)$ are constants in the 
present model, but they depend on details of flow dynamics (Wang et al. 2017).
We establish the 
$\OXYZ$ frame at the mass center of the binary black holes with mass $M_1$ and $M_2$ 
separated by $A_0$ (Fig. 1, right panel). The local frames of the primary and the 
secondary black holes are indicated, respectively, by $(\OXYZ)_1$ and $(\OXYZ)_2$. In $\OXYZ$, 
$\nn_{\rm obs}=(0,-\sin i_0,-\cos i_0)$ 
is the unit vector of observers toward to the center. $M_1$ has a distance of $A_1=qA_0/(1+q)$ 
to the center, where $q=M_2/M_1$. The Keplerian rotation velocity is
\begin{equation}
\Omega_0\!=\!%\sqrt{GM_1(1+q)/A_0^3}=
          5.4\times 10^{-9}\,M_8^{1/2}(1+q)^{1/2}A_{30}^{-3/2}\,{\rm s^{-1}},
\end{equation}
where $M_8=M_1/10^8\sunm$ and $A_{30}=A_0/30$ltd. Here the ``ltd'' is the distance that 
light travels in free space in one day.

\subsection{Composite transfer functions}
In order to obtain the projected velocity $g(\RR,v)$ for each BLR (I and II), we transform all velocities 
into their local frames. We move the detailed derivations of below formulations to Appendix B, C and D.
For the primary BLR-I, the observer's vector is given by rotating the $\OXYZ$ by $\pi+\phi_0$. 
For one cloud in BLR-I, it is located at $R_1(\cos \phi_{\rm B1},\sin\phi_{\rm B1},0$) and rotates with 
velocity $(-\sin\phi_{\rm B1}, \cos\phi_{\rm B1}, 0)V_{\rm K}$ in the orbital plane ($O_1-\theta_1\phi_1$), 
where $\theta_1$ and $\phi_1$ are the poloidal and toroidal angles. All transformations are tedious, but it is 
straightforward to obtain the total velocity $\vec{V}_1$ (see Supplementary Material for detailed derivations). 
We have the projected velocity of one cloud in 
the observer's frame as $V_{\rm d,1}^{\parallel}=\vec{n}_1\cdot\vec{V}_1$ explicitly as
\begin{eqnarray}\label{disk-v}
\frac{V_{\rm d,1}^{\parallel}}{V_{\rm K}}=
    \left[\left(\cos\theta_1 +\frac{\Omega_0 R_1}{V_{\rm K}}\right)q_1-\cos i_0\sin\theta_1\right]\cos\phi_{\rm B1}-
    \left(\!1 +\frac{\Omega_0 R_1}{V_{\rm K}}\cos\theta_1\!\right)\!q_2\sin\phi_{\rm B1}\nonumber \\
    -\frac{A_1\Omega_0}{V_{\rm K}}\cos\phi_0\sin i_0,
\end{eqnarray}
and the projected velocity function of $g_{\rm disk}=\delta\left(v-V_{\rm d,1}^{\parallel}\right)$.
The distance of light travel is 
$S_{\rm d,1}=R_1+\vec{r}_1\cdot\vec{n}_1=R_1(1 + q_2\cos\phi_{\rm B1}+q_3\sin\phi_{\rm B1})$,
where $q_1=\sin i_0 \cos(\phi_0+\phi_1)$, $q_2=\sin i_0 \sin(\phi_0+\phi_1)$,
$q_3= q_1 \cos\theta_1  - \cos i_0\sin\theta_1$ and $\vec{r}_1$ is the distance of the cloud to the mass center.
The effects of the orbital motion are an important parameter for the kinematics, which can be seen in Eq. (\ref{disk-v}).
In particular, each cloud has deviations from the primary orbital motion. However, the light travel 
remaines the same with a single AGN. In principle, this leads to kinematic features that distinguish binary BLRs 
from a single BLRs.

Given the emissivity law of $\epsilon_{\rm disk}=\epsilon_1\left(R/\Rin\right)^{-\gamma_1}$, 
we have the TF of the disk part of the BLR-I from Eq. (6) and after integrating over $R_1$ 
and $\phi_{\rm B1}$,
\begin{eqnarray}
\Psi_{\rm d,1}(v,t) = \frac{\epsilon_1}{4\pi}\int_{-\Theta_{\rm disk}}^{\Theta_{\rm disk}} \cos\theta_1d\theta_1
                          \int_{0}^{2\pi} d\phi_1
                          \sum_{i}\left(\frac{\Rin}{\Rdi}\right)^{\gamma} %\nonumber\\
                          \times \left(\frac{\Rdi}{ct}\right)
                          \left|\frac{dV_{\rm d,1}^{\parallel}}{d\phi_{\rm B1}}
                          \right|^{-1}_{\phi_{\rm B1}=\phi_{{\rm B1},i}^0},\quad
\end{eqnarray}
where $\Rdi=ct/\left(1+q_2\cos\phi_{\rm B1}+q_3\sin\phi_{\rm B1}\right)$ determined by $\phi_{{\rm B1},i}^0$, 
which is the $i-$th roots of the equation $v=V_{{\rm d,1}}^{\parallel}$.

For the case of inflows with only radial motion, the derivations are simpler than the disk case. The projected 
velocity is given by
\begin{eqnarray}
V_{\rm in,1}^{\parallel}=\Omega_0 R_1 q_1\sin\theta_1 
+V_{\rm in}\left(\cos i_0\cos\theta_1 -q_2\sin\theta_1\right)
-A_1\Omega_0\cos\phi_0\sin i_0,
\end{eqnarray}
and the projected velocity function  by
$g_{\rm in,out}=\delta(v-V_{\rm in,1}^{\parallel})$. The light travel distance from inflows is 
$S_{\rm in,1}= R_1\left(1 - \cos i_0\cos\theta_1 + q_2\sin\theta_1\right)$.
Given the emissivity law of $\epsilon_{\rm in,out}=\epsilon_{2,3}\left(R/\Rout\right)^{-\gamma_{2,3}}$ 
for inflows and outflows, respectively,  we have the 2D-TF of the inflow part
\begin{eqnarray}
\Psi_{\rm in,1}(v,t)=\frac{\epsilon_2}{4\pi}\int_{\pi/2-\Theta_{\rm flow}}^{\pi/2+\Theta_{\rm flow}} \sin\theta_1d\theta_1
                          \sum_i\left(\frac{\Rout}{\Rini}\right)^{-\gamma_2}%\nonumber\\
                          \times \left(\frac{\Rini}{ct}\right)
                          \left|\frac{dV_{\rm in,1}^{\parallel}}{d\phi_1}
                          \right|^{-1}_{\phi_{1,i}=\phi_{1,i}^0},\quad
\end{eqnarray}
where $\Rini=ct/\left(1-\cos i_0\cos\theta_1+q_2\sin\theta_1\right)$, $\phi_{1,i}^0$
is the $i-$th roots of the equation $v=V_{{\rm in,1}}^{\parallel}$. For outflows, the light travel 
difference is the same with the inflows ($S_{\rm out,1}=S_{\rm in,1}$), but the 
projected velocity differs. In such a case, we only need to change $V_{\rm in}$ into $V_{\rm out}$ for the 
2D-TF given by
\begin{eqnarray}
\Psi_{\rm out,1}(v,t)=\frac{\epsilon_3}{4\pi}\int_{\pi/2-\Theta_{\rm flow}}^{\pi/2+\Theta_{\rm flow}} \sin\theta_1d\theta_1
                          \sum_i\left(\frac{\Rout}{\Routi}\right)^{-\gamma_3}%\nonumber\\
                          \times\left(\frac{\Routi}{ct}\right)
                          \left|\frac{dV_{\rm out,1}^{\parallel}}{d\phi_1}
                          \right|^{-1}_{\phi_{1,i}=\phi_{1,i}^0},\quad
\end{eqnarray}
where $\Routi=ct/\left(1-\cos i_0\cos\theta_1+q_2\sin\theta_1\right)$, $\phi_{1,i}^0$ 
is the $i-$th roots of the equation $v=V_{{\rm out,1}}^{\parallel}$, and 
\begin{eqnarray}
V_{\rm out,1}^{\parallel}=\Omega_0 R_1 q_1\sin\theta_1 
+V_{\rm out}\left(\cos i_0\cos\theta_1 -q_2\sin\theta_1\right)
-A_1\Omega_0\cos\phi_0\sin i_0.\quad
\end{eqnarray}
There are only two integrations in Equation (11), and only one in (13) and (14), which 
are straightforward to perform. The derivations of $dV_{\rm d,1}^{\parallel}/d\phi_{\rm B1}$ 
and $dV_{\rm in, out,1}^{\parallel}/d\phi_1$ can be explicitly expressed in an extended 
form, but they are not difficult to calculate. Summing the three parts,
we have the entire 2D-TF of the BLR-I and -II, 
\begin{equation}
\Psi_i(v,t)=\Psi_{\rm d,i}(v,t)+\Psi_{\rm in,i}(v,t)+\Psi_{\rm out,i}(v,t),
\end{equation}
where $i=1,2$, respectviley.

The same can be done for the BLR-II according to the above formulations.
Summing all functions for each of the BLRs, we
then have the global 2D-TFs of the binary BLRs according to Eq. (\ref{psitot0}). 
The $(v,\tau)-$plane can be obtained by 
\begin{equation}
\tau=\frac{\int t\Psi_{\rm tot}(v,t)dt}{\int \Psi_{\rm tot}dt}.
\end{equation} 
Given the parameters of the binary BLRs, Eqs (5) and (17) will produce the kinematics of the
central regions, which can be compared with observations. Equation (18) shows delays with
velocity bins.

  \begin{figure}%\label{single_BLR}  
    \centering 
    \includegraphics[width=0.6\textheight]{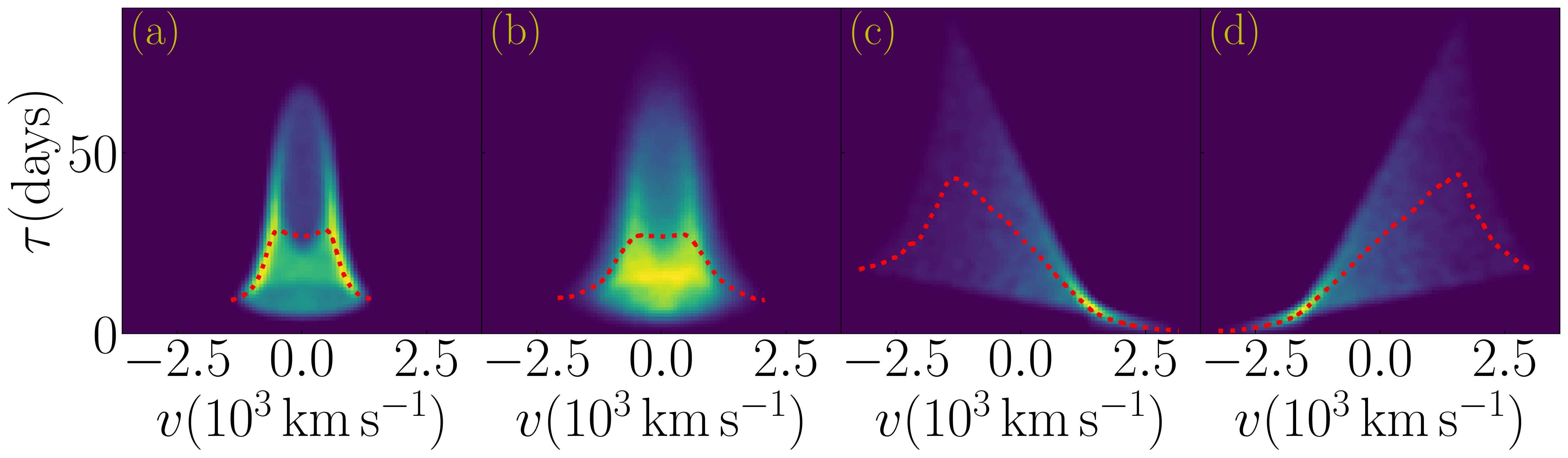} 
    \caption{\footnotesize 2D-TFs for a single BLR with different geometries, with red lines   corresponding to the $(v,\tau)$-planes. In panels (a) and (b): 
$\Theta_{\rm disk}=(5^{\circ}, 30^{\circ})$ represent geometrically thin and flattened disks; (c) and 
(d): $\Theta_{\rm flow}=45^{\circ}$ is for inflows/outflows. We took 
$(\alpha_0,\alpha;\beta_0,\beta;\gamma_{1,2,3})=(1.4,0;1.6,0.1;0.5)$, $\bhm=10^7\sunm$, 
$(\Rin,\Rout)=(9,45)$ltd and $i_0=30^{\circ}$. For a disk, we took $\Rmid=\Rout$, inflows and 
outflows ($\Rmid=\Rin$), respectively. For a single AGN, $(\Omega_0,A_0)=(0,0)$ are used in Eqs. 
(10, 12, 13).} 
  \end{figure}
  \label{single_BLR}

\section{Kinematic signatures}
For a single AGN, we set up the typical parameters of BLRs according to the $R-L$ relation for AGN
with sub-Eddington accretion rates (Kaspi et al. 2000; Bentz et al. 2013; Du et al. 2014, 2018). 
For a black hole with a mass of $\bhm=10^7\sunm$, the optical luminosity at 5100\AA\, is
$L_{5100}\approx \lambda_{\rm Edd}L_{\rm Bol}/\kappa_{\rm Bol}
         =1.4\times 10^{43}\,\lambda_{0.2}\kappa_{20}^{-1}\,M_7\,\ergs$ and the radius of the corresponding 
 BLR is $\langle R_{\rm BLR}\rangle\approx 12.5\,\lambda_{0.2}^{1/2}\kappa_{20}^{-1/2}M_7^{1/2}$ltd, where 
$L_{\rm Bol}$ is the bolometric luminosity, $\lambda_{\rm 0.2}=\lambda_{\rm Edd}/0.2$ is the Eddington ratio, 
and $\kappa_{20}=\kappa_{\rm Bol}/20$ is the bolometric correction factor (Jin et al. 2012). 
The inner and outer radii
follow $\Rin\!\!<\!\!\langle R_{\rm BLR}\rangle\!<\!\!\Rout$ ($\lambda_{\rm Edd}=0.2, M_7=1)$. 
Fig. 2 shows 2D-TFs. Geometrically thin disks have been extensively studied by BM82,
Robinson \& Perez (1990), Perez, et al. (1992), Welsh \& Horne (1991),  Chiang \& Murray (1996) and
Mangham, et al. (2017). The functions show
the 2D-TF as a symmetric bell-shape. The height and width of the bells are 
jointly determined by $(\Rin,i_0)$ and the width by $\Rin$. For a flattened disk BLR, the 2D-TF displays 
a fan-shape. Consisting of a series of thin disks with different orientations with respect to a remote 
observer, combinations of the individual bells make the fan-shape. Signatures of thin or flattened disks 
often appear among the reverberation-mapped AGNs (Bentz et al. 2010 for Arp 151; Grier et al. 2013
for Mrk 335, Mrk 1501, 3C 120). The inflows/outflows show a simple 
brush-shape as the response functions with 
slops, which is around those of the radial velocity profiles (follow the indexes of 
$\alpha$ and $\beta$). 

To illustrate the ionized gas kinematics in CB-SMBHs, we considered compositions 
of a wide range of possible BLR geometries. We took $M_1=10^8\sunm$, $q=0.5$, $\lambda_{\rm Edd}=10^{-2}$ 
(for both black holes), $A_0=30\,$ltd, and
the averaged radii of BLR sizes of (12.5, 8.4)\,ltd for the primary and secondary BLRs, respectively,
where $\kappa_{\rm Bol}=10$ (similar to NGC 5548 as the best-mapped
object (Li et al. 2016; Pancoast et al. 2014; Pei et al. 2017; Lu et al. 2016). 
We set $(\Rin,\Rout)_{1;2}=(7,15;\,4,10)\,$ltd.
We kept $R_{\rm out,1}+R_{\rm out,2}\lesssim A_0$ for a detached-BLR binary, 
and $(\Rin,\Rout)$ follow $\Rin\!\!<\!\!\langle R_{\rm BLR}\rangle\!\!<\!\!\Rout$ 
in each BLR. In such a binary, the maximum shifts are $V_{1,2}\approx (1690,3380)\,\kms$, 
implying an offset of about $4000\ \kms$ between the primary and the secondary. 

\begin{figure} 
    \centering 
    \includegraphics[width=0.75\textheight]{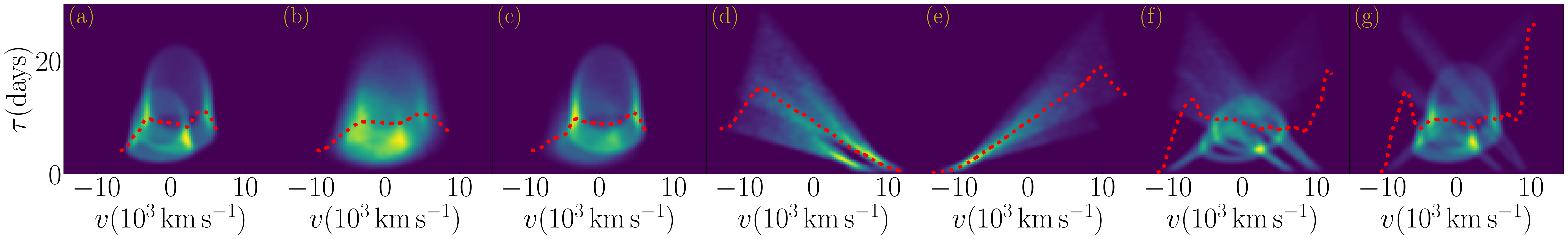} 
    \caption{\footnotesize
2D-TFs for compositions of same kinds of BLR geometries and the corresponding ($v,\tau$)-plane
(red lines) with $(i_0,\phi_0)=(30^{\circ},150^{\circ})$. ($\epsilon_{1,2,3}$; $\gamma_{1,2,3};
\alpha_0,\alpha;\beta_0,\beta)$ are the same with a single BLR. Panel 
{\it a} is for 2$\otimes$thin-disks; 
{\it b}: 2$\otimes$thick-disks; 
{\it c}: (thin$\otimes$thick)-disks;
{\it d}: 2$\otimes$inflows; 
{\it e}: 2$\otimes$outflows; 
{\it f}: 2$\otimes$(thin-disk+inflows+outflows) with smaller $\Rmid$, but
{\it g}: with larger $\Rmid$. Generally, they are
different from a single AGN. Here the symbol $\otimes$ represents a composition of one binary system.  } 
\end{figure} 

\begin{deluxetable}{lcccccccc}
\tablecolumns{9}
\tablewidth{0pc}
\tablecaption{Values of basic parameters of the model}
\tabletypesize{\footnotesize}
\tablehead{             &
\colhead{$M_{\bullet}$} &
\colhead{$R_1$}         &
\colhead{$R_0$}         &
\colhead{$R_2$}         & 
\colhead{$f_{\rm disk}$}&
\colhead{$f_{\rm in}$}  &
\colhead{$f_{\rm out}$} & 
\colhead{$\Theta_{\rm disk}$}\\
                &
($M_{\odot})$   & 
(ltd) & 
(ltd) &
(ltd) &
      &
      &
      &
($^{\circ}$)  
}
\startdata
Primary   &$10^8$&7  & 13  & 15 & 0.7 &  0.2 &   0.1 & 30\\
Secondary &--    &4  &  8  & 10 & 0.7 &  0.2 &   0.1 & 30\\ 
\enddata
\vglue -0.5cm
\tablecomments{These typical values are justified by NGC 5548. We assume Eddington ratio 
$\lambda_{\rm Edd}=0.01$. BLR's parameters follow the MCMC results of Pancoast et al. (2014).
$A_0$ is based on estimations in Li et al. (2016).
}
\end{deluxetable}

We took $\Psi_{\rm tot}=0.6\Psi_{\rm d,1}+0.4\Psi_{\rm d,2}$ ($\Gamma_0=2/3$ is assumed) for 
Fig. 3({\it a-c}); and $\Psi_{\rm tot}=0.6\Psi_{\rm in/out,1}+0.4\Psi_{\rm in/out,2}$ for panels 
({\it d,\,e}). Fig. 3{\it a} shows double-bells with an offset of 
$(V_1+V_2)\sin i_0\sin\phi_0\sim 1000\ \kms$ for a given $i_0$ and $\phi_0$,
providing observational signatures of CB-SMBHs in the 2D-TF as well as 
in the ($v,\tau)$-plane. Fig. 3{\it b} shows a
composition of 2$\otimes$thick-disks, the total 2D-TF is broadened significantly by the orbital
motion, and generally loses symmetry of the fans because of their offset velocity 
unless $\Gamma_0=1$. 
If $\Rin$ is very different in each BLRs, the asymmetry will be enhanced since the lag difference
follows $\Rin$. The flattened part of the  $(v,\tau)$-plane generally shows double peaks governed by
$i_0$. The 2D-TF of the (thin$\otimes$thick)-disks shown in Fig. 3{\it c} displays a hybrid
composite configuration of a bell and a fan.
2D-TFs of $2\otimes$inflows/outflows show relatively simple behavior in Fig. 3({\it d,e}),
where two mismatched brushes are controlled by orbital motion [brush lengths depend on $(\Rin,\Rout)$].
We took $R_{\rm mid,1;2}=(9,6;13,8)$ltd and $\Psi_{\rm tot}=0.6\Psi_1+0.4\Psi_2$ in Fig. 3{\it f} 
and 3{\it g}, and show the composite 2D-TFs of 2$\otimes$(thin-disk+inflows+outflows)-BLR
to be hybrid configurations influenced by $\Rmid$. The flux ratios of disk:inflow:outflow are 
(0.3:0.5:0.2; 0.7:0.2:0.1) for BLR-I, and -II, respectively.

Fig. 4 panels show composite 2D-TF configurations of 
$\Psi_{\rm tot}=0.6\Psi_{\rm d}+0.4\Psi_{\rm in,out}$, where $\Psi_{\rm d}$, $\Psi_{\rm in,out}$ 
are 2D-TFs of one single disk, inflow or outflow, respectively. 
The thin-disk$\otimes$(inflows, outflows) are easily distinguished as well as inflows$\otimes$outflows, 
but the thick-disk$\otimes$(inflows,outflows) are complicated depending several factors
$\Rin$ and $(i_0,\phi_0)$. The most complicated configuration 
is the 2$\otimes$(thick-disk+inflows+outflows), which is additionally affected by $\Rmid$. The flux
ratios are 0.6:0.3:0.2 and 0.85:0.15:0.05 in Fig. 4({\it f,g}), respectively.
From these configurations in Fig. 3 and 4, we find that CB-SMBHs with 
disk$\otimes$inflows/outflows, 2$\otimes$thin-disks and inflows$\otimes$outflows 
are relatively easy to distinguish from others.

  \begin{figure} 
    \centering 
    \includegraphics[width=0.75\textheight]{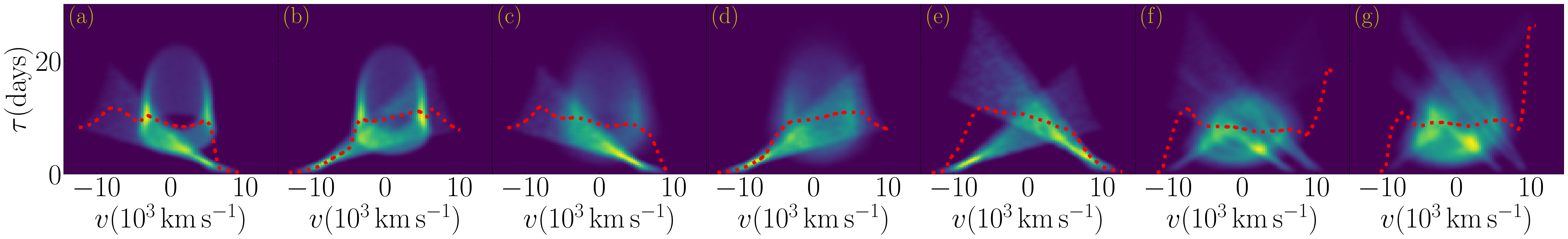} 
    \caption{\footnotesize 
2D-TF for compositions of different kinds of BLR geometries and corresponding $(v,\tau)$-planes
(red lines). The parameters are the same with Fig.3. Panel 
{\it a} is for thin-disk$\otimes$inflows; 
{\it b}: thin-disk$\otimes$outflows; 
{\it c}: thick-disk$\otimes$inflows;
{\it d}: thick-disk$\otimes$outflows;
{\it e}: inflows$\otimes$outflows;
{\it f}: 2$\otimes$(thick-disk+inflows+outflows) with small $\Rmid$, but
{\it g} with large $\Rmid$.
They are significantly different from a single AGN.} 
  \end{figure}

We show more 2D transfer functions for different parameters.
The fixed parameters are listed in Table 1. Here $f_{\rm disk}$, $f_{\rm in}$ and $f_{\rm out}$
are fractions of disk, inform and outflow components to the total.
In Fig. 5, we take $i_0 = 10^{\circ}, 20^{\circ}, 30^{\circ}, 40^{\circ}, 50^{\circ}$ for panel 
{\it a,b,c,d} for $A_0=30$ltd, $\phi_0=150^{\circ}$ and $q=0.5$, respectively.
In Fig. 6, we adjust $A_0$ for the 2D-TFs with fixed $i_0=30^{\circ}$, $\phi_0=150^{\circ}$
and $q=0.5$, where we take $A_0=30,40, 50,60$ltd for panel {\it a,b,c,d}, respectively.
In Fig. 7, we show 2D-TFs for various mass ratios of $q=0.1, 0.3, 0.5, 0.7, 0.9$, and
$A_0=30$ltd, $\phi_0=150^{\circ}$ and $i_0=30^{\circ}$. Dependence of the 2D-TF on parameters
can be seen from Fig. 5-7.

In principle, the parameters among $A_0$ and ($R_{\rm out,1},R_{\rm out,2}$) can be arbitrarily 
chosen (as well as Eddington ratios, black hole masses and their mass ratios). We restrict our 
parameter space with 
$A_0\ge (R_{\rm out,1}+R_{\rm out,2})$ as the valid range of the present model. If the binary BLRs
merger, the common envelope-BLR will be formed first, and then totally merge as a unified BLR, as 
pointed out, but the 2D-TF can still be calculated in principle.

Finally, we would like to point out that the orbital motions as the major features of 
CB-SMBHs can be distinguished from a single BLR provided that the individual BLRs have regular geometers.
More figures for larger ranges of parameters in Table 1 can be conveniently plotted by using the 
semi-analytical expressions in this paper, but in principle, the total 2D-TF is composite from two
individual 2D-TFs through the coupling coefficient $\Gw$. 
The expression makes it easy to compare with observational data by fitting 
in order to search for CB-SMBHs through RM-surveys.

\section{Summary and discussion}
As spatially unresolved sources, close binaries of supermassive black holes have been expected 
to be identified for detection of low-frequency gravitational waves.
In this paper, we demonstrate compositions of 2D-TFs of various kinds of geometries of binary 
BLRs. Composite 2D-TFs are generally distinguished from a single AGN due to the orbital motion 
of the binary, although some of them have complicated configurations. This provides a promising 
way of hunting CB-SMBHs in AGNs through RM-campaigns. Unlike stellar black hole binaries,
the CB-SMBH merger 
timescale of $t_{\bullet}\approx 2.4\times 10^4\,A_{30}^4M_8^{-3}q^{-1}(1+q)^{-1}$\,yrs due 
to GW radiation (Peters \& Mathews 1963) allows astronomers and physicist 
to set up GW physics and the orbital dynamics of CB-SMBHs to test general relativity.

Only about 15 AGNs or so have 2D-TFs due to RM-data quality, but the qualities are 
still not good enough to produce high-quality 2D-TFs in order to clearly reveal the kinematics 
features of orbital motions. It is hardly to justify the presence of 
the kinematic features of CB-SMBH. The inquiry of high-fidelity RM-campaigns for this goal
depends on several factors: 1) homogeneous and high cadence; 2) reasonable spectral resolution; 
3) spectral calibration should be improved for the shape changes of H$\beta$ profiles (\oiii\, 
is then a poor calibrator for this goal). Targets of future RM-campaigns should focus 
on double-peaked Seyfert galaxies (high-$i_0$), whose signals of orbital motion are 
more easily detected\footnote{Moreover, AGNs with geometrically thin BLRs are preferred 
because of more sharper features of 2D-TFs. Considering that BLR could 
follow the dusty torus (Wang et al. 2017), we should select those sources with weaker NIR emission
that have the smaller covering factors
of the torus as well as the thin BLR.}. Simulations will be carried out for inquiry of high-quality 
2D-TFs (e.g. Horne et al. 2004) as powerful tools to find CB-SMBHs. Actually, we started a long-term 
campaign of Monitoring AGNs with H$\beta$ Asymmetry (MAHA project) in 2016 through the Wyoming Infrared 
Observatory (WIRO) 2.3m telescope for CB-SMBHs.
 
The present features of orbital motion of binary systems commonly appear in the 2D-TFs, but
they will be totally distorted if the binary BLRs are chaotic. Future work to improve the 
present model can be done in two main aspects. Gaseous 
dynamics should be given in a self-consistent way, which involves radiation hydrodynamics, 
but this is very complicated even for the case given the BLR origin. Another aspect should deal with
photoionization of the gas around the accreting black holes, which is done (e.g. Waters et al. 
2016; Mangham et al. 2017), or some other models (e.g. Goad \& Korista 2015). The present work 
is to draw attention of astronomers that the orbital motion of the CB-SMBHs provides observable 
features. Numerical simulations of the BLRs in the merging binary black holes are highly desirable.

Finally, we would like to point out that there is a contact stage of CB-SMBHs 
(if $A_0\le R_{\rm out,1}+R_{\rm out,2}$) before their mergers: a common envelope-BLR 
is formed first, and then totally merger as an unified one, which is ionized by the binary accretion 
disks before binary black holes merger. The 2D-TFs are complicated than the present, but 
can be calculated in principle.

\acknowledgments{The authors are grateful to an anonymous referee for his/her 
useful report improving the presentation of the paper. 
We express thanks to the members of the IHEP-AGN group for helpful 
discussions. M. Brotherton is thanked for carefully reading of the manuscript.
This research is supported by National Key R\&D Program of China (grant
2016YFA0400701) and grants NSFC-11173023, -11133006, -11373024, -11233003 and -11473002, and
by CAS Key Research Program of Frontier Sciences, QYZDJ-SSW-SLH007.
}

\appendix
\section{About $\Gw$}
We give a brief discussion about $\Gw$ and its effects. In order to understand this effect, 
we expand 
$\Gw=\Gamma_0+\Gw^{\prime}$ as a perturbation,
where $|\Gw^{\prime}|\ll \Gamma_0$. Inserting into Eq. (\ref{psitot}), we have
\begin{equation}
\Psi_{\rm tot}(v,t)=\Psi_{\rm tot}^0(v,t)+\left(\Delta \Psi_2-\Delta\Psi_1\right)/(\Gamma_0+1)^2,
\end{equation}
where $\Psi_{\rm tot}^0(v,t)$ is given by Eq. (5), and
\begin{equation}
\Delta\Psi_k=(2\pi)^{-1}\int d\vec{R}_k{\cal{H}}_k
                    \int_{-\infty}^{+\infty}\!\!d\omega\Gw^{\prime}e^{-i\omega t_k},\quad\, (k=1,2).
\end{equation}
This implies that the variation-coupling effects in the composition of $\Psi_{1,2}$ can be cancelled 
in some degree, improving the approximation of Eq. (\ref{psitot0}) for a binary system with minor difference
between the binary BLRs. 

For big different cases, we have to consider the couple effects in details.
The PDS is assumed to be given generally in a form of 
$
P(\omega)=p_0/\left[1+\left(\omega/\omega_0\right)^{\gamma_{\omega}}\right],
$
where $\omega_0$ is the break frequency and $p_0$ is a constant  (e.g. Kelly et al. 2009). 
Considering $\tilde{L}_{\rm c}\propto P(\omega)^{1/2}$, we have
\begin{equation}
\Gw=\Gamma_0\left[\frac{1+\left(\omega/\omega_2\right)^{\gamma_2}}{1+\left(\omega/\omega_1\right)^{\gamma_1}}\right]^{1/2},
\end{equation}
where $\Gamma_0=(p_1/p_2)^{1/2}$. In order to analytically discuss the $\Gw-$effects, we approximate
\begin{equation}
\Gw\approx \left\{\begin{array}{ll}
\Gamma_0 & {\rm for}~\omega\lesssim(\omega_1,\omega_2),\\
         &     \\
\Gamma_0^{\prime}\omega^{\alpha_0}   & {\rm for}~\omega\gtrsim(\omega_1,\omega_2),
\end{array}\right.
\end{equation}
where
$\Gamma_0^{\prime}=\Gamma_0\left(\omega_1^{\gamma_1}/\omega_2^{\gamma_2}\right)^{1/2}$ and 
$\alpha_0=(\gamma_2-\gamma_1)/2$. The second term shows the effects of the break frequency 
and non-random walk variations
($\gamma_1\neq\gamma_2\neq2$).

In principle, $(\omega_{1,2},\gamma_{1,2}, \Gamma_0)$ are fully unknown with a prior in a binary system. 
In this Appendix, we only show the influence of $\Gw$. For $\alpha_0=0$, it returns to the simplest case
with $\Gw=\Gamma_0$. For a general discussion, we have
\begin{equation}
{{\cal{Q}}_{1,2}(v,t,R_{1,2})}=\frac{1}{2\pi}\int_{-\infty}^{+\infty}\!\!d\omega
      \left[\frac{e^{-i\omega t_1}}{1+\Gamma_0^{\prime}\omega^{\alpha_0}},
      \frac{e^{-i\omega t_2}}{1+\left(\Gamma_0^{\prime}\right)^{-1}\omega^{-\alpha_0}}\right].
\end{equation}
The $\Gw$ as a function of $\omega$ makes the response be broadened generally. In order to clarify the 
$\Gw$-effects, we take $\alpha_0=2$ for an analytical discussion, 
\begin{equation}
{{\cal{Q}}_1(v,t,R_1)}=\frac{1}{4\sqrt{\pi\Gamma_0^{\prime}}}e^{-|t_1|/\sqrt{\Gamma_0^{\prime}}};\quad
{{\cal{Q}}_2(v,t,R_2)}=\delta(t_2)-\frac{1}{4\sqrt{\pi\Gamma_0^{\prime}}}e^{-|t_2|/\sqrt{\Gamma_0^{\prime}}}.
\end{equation}
It can be easily seen that $\Gw$-effects broaden the function ${\cal Q}_{1,2}$, and thus the 2D-TF.
This effect mixes with the spatial distributions of emissivity ($\epsilon_{1,2}$), however, this cannot
change effects of the orbital motion of the binary.  In future practice, it could be reasonable to 
take $\Gw$ as a free parameter in fitting the observed 2D-TFs obtained from the high-fedility
RM data.

\section{Coordinate transformation}
In this supplementary material, we provide detailed derivations of equations (9-14) and
more plots for different parameters of close binary supermassive black holes.
All the derivations involve the transformation of coordinates from different frames. For the primary
black hole, the distance to the mass center is $A_1$, and
the transformation of coordinates from $O-XYZ$ to $(O-XYZ)_1$ is
\begin{equation}
    \begin{pmatrix}
        x_1\\
        y_1\\
        z_1
    \end{pmatrix}
    =
    \begin{pmatrix}
        -\cos\phi_0 & -\sin\phi_0 & 0 \\
        \sin\phi_0 & -\cos\phi_0 & 0 \\
        0 & 0 & 1
    \end{pmatrix}
    \begin{pmatrix}
        x\\
        y\\
        z
    \end{pmatrix}
    +
    \begin{pmatrix}
        A_1\\
        0\\
        0
    \end{pmatrix}.
\end{equation}
Similarly,  transformation of coordinates from  $(O-XYZ)_1$ to $O-XYZ$ is
\begin{equation}
    \begin{pmatrix}
        x\\
        y\\
        z
    \end{pmatrix}
    =
    \begin{pmatrix}
        -\cos\phi_0 & \sin\phi_0 & 0 \\
        -\sin\phi_0 & -\cos\phi_0 & 0 \\
        0 & 0 & 1
    \end{pmatrix}
%    \cdot
    \begin{pmatrix}
        x_1 - A_1\\
        y_1\\
        z
    \end{pmatrix}.
\end{equation}
Integrations with the $\delta$-function can be performed as
\begin{equation}
\int f(x)\delta[g(x)] dx=\sum_{i=1}^n\frac{f(x_i)}{|g^{\prime}(x_i)|},
\end{equation}
where $x_i$ is the $i$-th real root of $g(x)=0$ and $g^{\prime}(x_i)=dg(x)/dx|_{x=x_i}$.
It is useful to simplify the 2D transfer functions.

\section{Transfer function of the disk part}
We provide formulas for the projected velocities of BLR-I, and BLR-II can be
obtained in a similar way. For a cloud in the plane $O_1\!-\!X_1Y_1$, 
its coordinates are 
$(\cos\phi_{\rm B1}, \sin\phi_{\rm B1}, 0)R_{1}$ and the velocity is 
$(-\sin\phi_{\rm B1}, \cos\phi_{\rm B1}, 0)V_{\rm K}$. Considering that we rotate the cloud 
onto the plane $O_1-X_{\rm c1}Y_{\rm c1}$ with angles of $(\theta_1,\phi_1$), we have
\begin{equation}
    \vec{r}_1 = R_1    
    \begin{pmatrix}
        \cos\phi_1 & -\sin\phi_1\cos\theta_1 &  \sin\phi_1\sin\theta_1 \\
        \sin\phi_1 &  \cos\phi_1\cos\theta_1 & -\cos\phi_1\sin\theta_1 \\
        0 & \sin\theta_1 & \cos\theta_1
    \end{pmatrix}
    \begin{pmatrix}
        \cos\phi_{\rm B1}\\
        \sin\phi_{\rm B1}\\
        0
    \end{pmatrix},
\end{equation}
and
\begin{equation}
    \vec{v}_1 =V_{\rm K}
    \begin{pmatrix}
        \cos\phi_1 & -\sin\phi_1\cos\theta_1 &  \sin\phi_1\sin\theta_1 \\
        \sin\phi_1 &  \cos\phi_1\cos\theta_1 & -\cos\phi_1\sin\theta_1 \\
        0 & \sin\theta_1 & \cos\theta_1       
    \end{pmatrix}
    \begin{pmatrix}
        -\sin\phi_{\rm B1}\\
        \cos\phi_{\rm B1}\\
        0
    \end{pmatrix}.
\end{equation}
The coordinates of the cloud in the $O-XYZ$ frame can be expressed by equivalent coordinates in the $(O-XYZ)_1$ frame
\begin{equation}
    \vec{r}_0 =     
    \begin{pmatrix}
        -\cos\phi_0 & \sin\phi_0 & 0 \\
        -\sin\phi_0 & -\cos\phi_0 & 0 \\
        0 & 0 & 1
    \end{pmatrix}
%    \cdot
    (\vec{r}_1 - \vec{A}_1),
\end{equation}
where $\vec{A}_1 = (A_1,0,0)$. The rotational velocity of the cloud co-rotating with
the $O-XYZ$ frame is
\begin{equation}
    \vec{v}_{\rm r0} = \vec{\Omega}_0 \times \vec{r}_0,
\end{equation}
where $\vec{\Omega}_0 = (0,0,\Omega_0)$. The rotational velocity of the cloud in the $(O-XYZ)_1$ 
frame is
\begin{equation}
    \vec{v}_{\rm r1} =
    \begin{pmatrix}
        -\cos\phi_0 & -\sin\phi_0 & 0 \\
        \sin\phi_0 & -\cos\phi_0 & 0 \\
        0 & 0 & 1
    \end{pmatrix}
    \vec{v}_{\rm r0}.
\end{equation}
Considering the direction of the line of sight in the $(O-XYZ)_1$ frame is given by 
\begin{equation}
    \vec{n}_1 = 
    \begin{pmatrix}
        -\cos\phi_0 & -\sin\phi_0 & 0 \\
        \sin\phi_0 & -\cos\phi_0 & 0 \\
        0 & 0 & 1
    \end{pmatrix}
%    \cdot
    \begin{pmatrix}
        0\\
        -\sin i_0\\
        -\cos i_0 \\
    \end{pmatrix},
\end{equation}
we have the velocity of the cloud projected on the direction $\vec{n}_1$
\begin{eqnarray}
\vec{n}\cdot \vec{V}_1&=& \vec{n}_1\cdot(\vec{v}_{\rm r1} + \vec{v}_{1})\nonumber\\
      &=&-A_1\Omega\cos\phi_0\sin i_0+ 
       \left[-V_{\rm K}\cos i_0\sin\theta_1 + (V_{\rm K}\cos\theta_1 +\Omega R_{\rm c1} )
       \sin i_0 \cos(\phi_0+\phi_1)\right]\cos\phi_{\rm B1}+ \nonumber \\
    & &[-(V_{\rm K} + \Omega R_{\rm c1}\cos\theta_1)\sin i_0 \sin(\phi_0+\phi_1)]\sin\phi_{\rm B1}.
\end{eqnarray}
The time lag of the line emission is
\begin{equation}
    R_1 + \vec{r}_1\cdot\vec{n}_1 = R_1\{ 1 + \sin i_0 \sin(\phi_0+\phi_1)\cos\phi_{\rm B1} + [\sin i_0 \cos\theta_1 \cos(\phi_0+\phi_1) - \cos i_0\sin\theta_1 ]\sin\phi_{\rm B1} \}.
\end{equation}
The transfer function of the disk is
\begin{equation}
\Psi_{\rm d,1}(v,t) = \frac{\epsilon_1}{4\pi} \int_{R_{\rm d,1}}^{R_0}  
                       \left(\frac{R_1}{R_{\rm d,1}}\right)^{-\gamma_1} d R_1 
                       \int_{-\Theta_{\rm disk}}^{\Theta_{\rm disk}} \cos\theta_1 d\theta_1 
                       \int_0^{2\pi} d\phi_1 \int_0^{2\pi}d\phi_{\rm B1} \delta(X_1)\delta(X_2),
\end{equation}
where $X_1 = v -  \vec{n}_1\cdot(\vec{v}_{\rm r1} + \vec{v}_{1})$ and $X_2 = ct - R_1- \vec{r}_1\cdot\vec{n}_1$.
Equation (11) can be obtained by integrating $X_1$ and $X_2$. 

\section{Transfer function of inflow/outflow part}
The velocity fields of inflows or outflows are given by Eq. (6).
Coordinates of one cloud in $(O-XYZ)_1$ frame are
\begin{equation}
    \vec{r}_{\rm 1} = R_1 
    \begin{pmatrix}
        \sin\theta_1\cos\phi_1\\
        \sin\theta_1\sin\phi_1\\
        \cos\theta_1
    \end{pmatrix},
\end{equation}
and its velocity is
\begin{equation}
    \vec{v}_{\rm in,1} = -V_{\rm in, 1}
    \begin{pmatrix}
        \sin\theta_1\cos\phi_1\\
        \sin\theta_1\sin\phi_1\\
        \cos\theta_1
    \end{pmatrix}.
\end{equation}
Similarly, we have the projected velocity as
\begin{eqnarray}
%\vec{n}_1\cdot\vec{V}_{\rm in,1}&=&
\vec{n}_1\cdot(\vec{v}_{\rm r1} + \vec{v}_{\rm in,1})=% \nonumber\\
 -A_1\Omega\cos\phi_0\sin i_0 + \Omega R \cos(\phi_0+\phi_1)\sin\theta_1\sin i_0+ \nonumber \\
    V_{\rm in,1}[\cos i_0\cos\theta_1 -\sin(\phi_0+\phi_1)\sin\theta_1\sin i_0],
\end{eqnarray}
and
\begin{equation}
    R_1+ \vec{r}_1\cdot\vec{n}_1 = R_1[1 - \cos i_0\cos\theta_1 + \sin(\phi_0+\phi_1)\sin\theta_1\sin i_0 ].
\end{equation}
The transfer function of the inflow part is
\begin{equation}
\Psi_{\rm in,1}(v,t) = \frac{\epsilon_2}{4\pi} 
                       \int_{R_{\rm in,1}}^{R_{\rm out}}  \left(\frac{R_1}{R_{\rm in,1}}\right)^{-\gamma_2} d R_1 
                       \int_{\pi/2-\Theta_{\rm disk}}^{\pi/2+\Theta_{\rm disk}} \sin\theta_1 d\theta_1 
                       \int_0^{2\pi} d\phi_1 \delta(Y_1)\delta(Y_2),
\end{equation}
where $Y_1=v-\vec{n}_1\cdot(\vec{v}_{\rm r1} + \vec{v}_{\rm in,1})$ and 
$Y_2 = ct - R_1- \vec{r}_1\cdot\vec{n}_1$. Integrations over $Y_1$ and $Y_2$ yield Equation (13).
Similar derivations can be done for outflows as shown in Equation (14).

\clearpage

  \begin{figure} 
    \centering 
    \includegraphics[width=0.8\textheight]{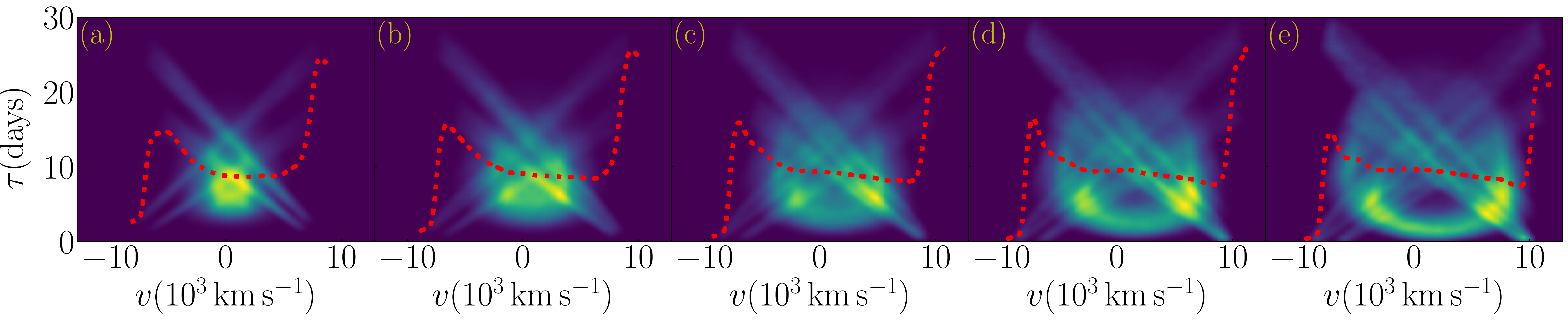} 
    \caption{\footnotesize The 2D-TF of binary black holes, showing dependence on inclinations 
of $i_0$, with different parameters given by Table 1.} 
  \end{figure} 
  \begin{figure} 
    \centering 
    \includegraphics[width=0.8\textheight]{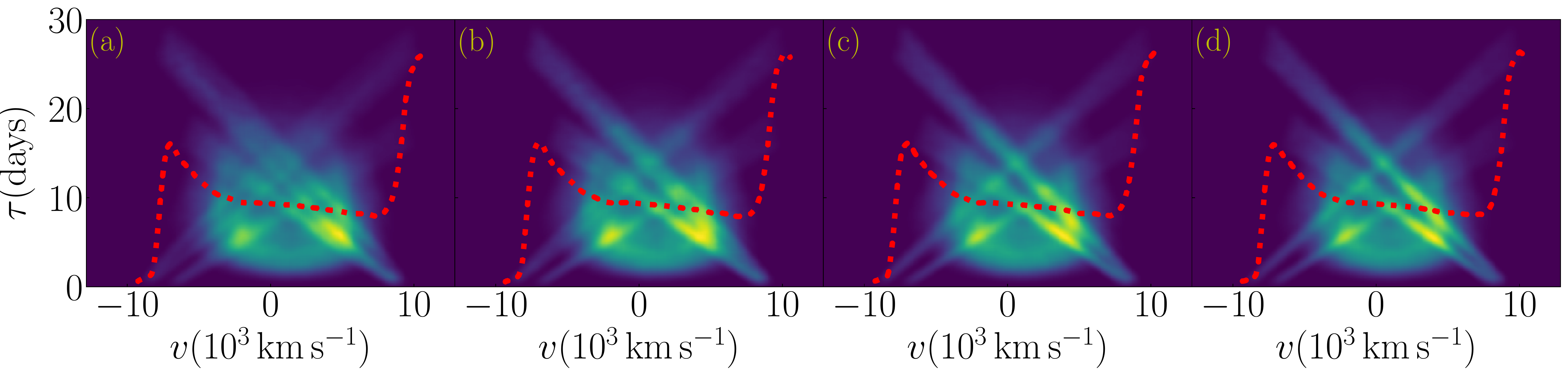} 
    \caption{\footnotesize The 2D-TF of binary black holes, showing dependence on binary 
separations of $A_0$, with different parameters given by Table 1.} 
  \end{figure} 
  \begin{figure} 
    \centering 
    \includegraphics[width=0.8\textheight]{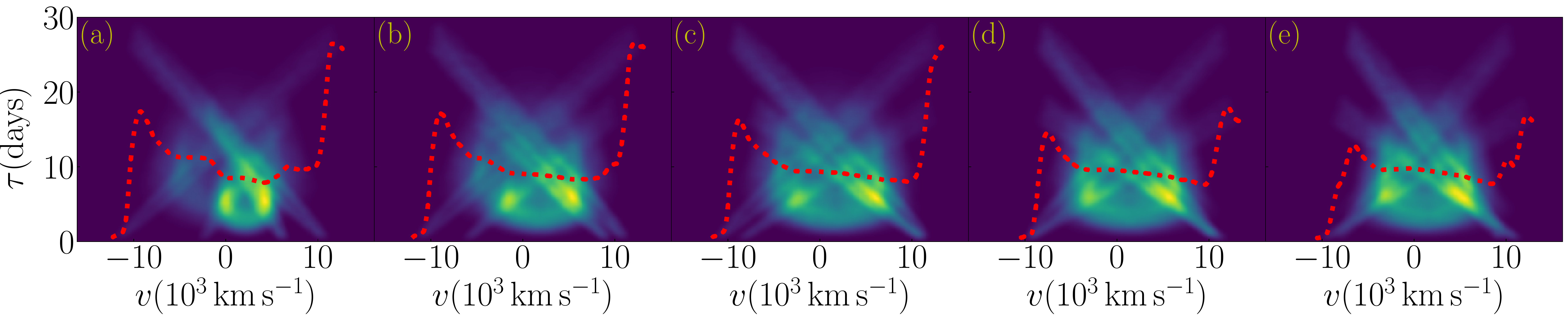} 
    \caption{\footnotesize The 2D-TF of binary black holes, showing dependence on mass ratios, 
with different parameters given by Table 1.} 
  \end{figure} 
\end{document}